# WiFiCue: Public Wireless Access Security Assessment Tool

Author: Jonathan K Adams, jka40138@marymount.edu
Advisor: Sally Vandeven
Accepted: September 30th 2019

## Abstract

Public wireless access points are commonly provided by governments, businesses, schools and other organizations and provide access to the Internet for numerous use cases and can present varying degrees of risk to users. While there are steps that can be taken to mitigate public Wi-Fi risks, ranging from avoidance to the application of end-to-end encryption, application specific encryption, and other technologies and tools, these options are not always viable. This paper examines risks associated with Wi-Fi from on a network-by-network perspective. Recommender Systems are presented as part of a proposed mechanism for informing users of the risks of connecting to a specific access point. Implementing prototype architecture for this purpose is examined.

# 1. Introduction

Public wireless access points present an untrusted computing environment for users, whether they are witting of the threat or not. One of the remedies often suggested is to utilize a Virtual Private Network [VPN]. On some wireless networks, VPNs may be disallowed. Even if allowed, Virtual Private Networks vary greatly in their security, and depending upon the vendor, may introduce additional risks and reasons for concern. Additionally, depending upon the application to be used, VPNs can inject network latency that makes some uses suffer, resulting in a poor user experience.

Some key factors in assessing the security of a public wireless network are what version of 802.11 is implemented, whether the wireless access point provides the option for encryption, and if so, what the level of encryption is and what mechanism is used for key exchange. Other less obvious factors include the overall security of the network the wireless access point is plugged into, the vendor and operating system or firmware version and patch level of the device(s) user data will transit, as well as the physical security and accessibility of the access point or access points on the network, etc. Initial tool for this effort development is on the Android platform, however once a fully functional capability is built, other platforms will be investigated as well. In short, this tool aims to address an observed capability gap resulting in no current straightforward way for an end user to quickly gain enough information to be informed about the security of the environment and risk to their data and privacy.

Recommender systems can potentially mitigate aspects of this threat. Recommender systems are most commonly, if not exclusively, utilized to rate the service consumed by customers. Examples of this include the website Yelp, which allows restaurant patrons to rate the service they have received, the quality of the food, and various other factors that comprise the customer experience when ordering out or visiting a restaurant. In practice, such a system could be utilized to capture the quality of service, particularly the factors that are known to impact security of a given Wi-Fi access point or network. However, there are limitations and assumptions that underlie this approach. The approach is not a solution that fully eliminates risk associated with utilizing publicly provided Wi-Fi. This concept does, however, have significant upside in helping users to avoid negative consequences of connecting their devices to high-risk Wi-Fi access points and networks based on a set of defined criteria and thresholds.

# 2. The State of 802.11 Wireless Security

A generally accepted position is that Wi-Fi networks present a series of security challenges that are not present with wired networks. These threats are partially tied to the media type – with over- the- air data being physically more difficult to secure than wired media, especially fiber. On a well-engineered device in wired ethernet environments, it is possible to completely eliminate visibility to traffic, even if a device is plugged in. In Wi-Fi environments, radio frequency waves are transmitted without device-based controls and can be passively captured and analyzed. These threats are also tied to protocol issues and the integration that is required to deliver Layer 1 and Layer 2 connectivity. Once the traffic is captured, it is possible to dissect and attack the protocol, cryptography, and access-related weaknesses across the various versions of the 802.11 protocol.

## 2.1.  Geo-Fencing

Tackling the physical aspect of controlling Wi-Fi coverage, Sheth, Seshan and Wehterall (2009) proposed a unique solution known as Geo-Fencing that relies on specialized directional antennae and tight control and thresholds for transmit-receive power and direction. Geo-Fencing is an effective mechanism for physically controlling Wi-Fi networks. A notable limitation in Geo-Fencing is that a requirement for fairly precise integration between the physical security measures and the technical implementation solution (Rahimi, Maimaiti, and Zincir-Heywood, 2014). On the surface and from a cost-benefit perspective, this does not seem to be a realistic solution for most use cases that surround publicly provided Wi-Fi. Most applications involved with this technology are for marketing or limiting access to networks with sensitive data on them, not mitigating the physical vulnerabilities of public Wi-Fi. Additionally, the specialized hardware and pre-implementation analysis would significantly increase the cost of implementing a public Wi-Fi and is not likely to be considered worth the expenditure if only used to provide public Internet access.

## 2.2.   802.11 Open Networks

In regard to Layer 2 security, the various iterations of 802.11 have some significant shortcomings. If a network access point is configured in "Open" mode, without encryption, it has to be considered an untrusted environment. On Open Wi-Fi networks network traffic is transmitted in without any of the available 802.11 encryption methods enabled. The encryption level of the traffic is dependent upon higher layer protocol capabilities such as Transport Layer Security, to protect data. When using an unsecured access point, not only are the network and the endpoints in it vulnerable to man-in-the-middle and denial of service attacks, there is also no expectation of security for the data transiting the network. Data can be captured and analyzed passively, assuming that it is not encrypted at a high-level. Sepehrdad et al. (2013) describe WiFi as vulnerable to passive scanning at long distances where Intrusion Detection System detection is minimal. From a user perspective, best practices for using public Wi-Fi suggest using a secure endpoint host policy, such as a firewall denying all inbound traffic, minimize the exposure of services on the computer (Goldborough, 2015). Cheng et al. (2013) recommend utilizing Virtual Private Networking [VPN] to provide robust encryption of traffic. This also has limitations and can negative affect the users experience through additional delay due to traffic latency.

## 2.3.   WEP and WPA

Wired Equivalent Protection [WEP] provides an extremely limited pretense of security. Data in transit is encrypted, however the encryption is not cryptophucally strong and can be broken. The protocols handshake is also insecure. Decrypting captured WEP traffic requires tools but doesn't require a high level of skill or specialized equipment and can be accomplished fairly quickly under certain circumstances. As noted, the weaknesses of WEP are not limited to the encryption strength. The WEP protocol itself is vulnerable. Vihbuti (2005) describes a weakness in the initialization vector as well as the problematic use of the same key for authentication and encryption. Both of these issues are significant and contribute to WEP being retired by the Wi-Fi Alliance in 2014 (Wi-Fi Alliance, 2019). Despite being retired, there are many access points deployed that are capable of implementing WEP, although most users have moved to newer protocols. Very similar usage practices should govern utilizing a public access point secured with WEP, as would an Open access point.

The next iteration of Wi-Fi security, Wi-Fi Protected Access [WPA], particularly when using Temporal Key Integrity Protocol [TKIP], is no longer considered secure, having been deprecated (Adnan et al., 2015). Sheldon (2002) describes that a number of protocol and protocol implementations to the initial version of WPA resolved some of the weaknesses in WEP, to include a larger key size. However, the protocol itself retained weaknesses as well as brought in new vulnerabilities and is also not considered secure. Vanhoef and Piessens (2013) demonstrate attacks on Temporal Key Integrity Protocol [TKIP], which is a key exchange protocol utilized to enable WPA to run on WEP compliant hardware, resulting in successful application of both a denial of service attack as well as decryption attacks. These are not an exhaustive list of security shortcomings that are executable and affect WPA users. As noted by Wong (2003), WPA was intended to be a temporary solution that implements aspects of 802.11i. It was available via software or firmware upgrade and did not require new access point hardware. This trade-off plays directly into the lack of robust security in the implementation.

WEP was an inherently insecure design, and WPA, while improved, still suffers from a number of security vulnerabilities, including incorporating the flawed Wi-Fi Protected Setup [WPS]. Bongard (2014) and Sanatinia, Narain, and Noubir (2013) identify issues in WPS including security that is vulnerable to brute-force attack enabling quick compromise and reconfiguration of access points, and subsequently, clients connected to them. While not particularly of concern on public open access points, WPS should not be enabled.  From broader perspective, WPA, as noted by Sari and Karay (2015), is able to use an insecure cryptography algorithm (RC4) and is subject to brute-force attack and denial of service attacks.

## 2.4.   WPA2

Wireless Protected Access version 2 [WPA2] addresses many of the shortcomings of WEP and WPA, implementing a more robust cryptography algorithm. Lashkari, Danhesh, and Samadi (2008) note that WPA2 fully implements 802.11i. IEEE 802.11i addresses many security problems involved with WEP, enables backward compatibility and improves protections for public Wireless Local Area Networks, without. focusing specifically on them (Chaplin et Al, 2005). Due to the encryption strength, brute-force is possible but may not be timely if proper password complexity rules are followed. Utilizing EAP largely mitigated the threat of dictionary attacks initially. However, eventually vulnerabilities were discovered. The WPA Enterprise configuration is more robust, but as noted by Sari and Karay (2015) is no trivial to implement. Robyns et al. (2014) demonstrates a practical attack that worked in a specific scenario against Apple devices, one of multiple Man-In-The-Middle [MITM] or MITM-like attacks. EAP utilizing TLS with public/private keys bypasses the vulnerabilities associated with most other EAP authentication protocols. Bartoli et al (2018) demonstrate an attack that works with varying degrees of success against varying OS platforms.

## 2.5.   WPA3 and Beyond

In 2018, Wireless Protected Access Version 3 [WPA3] was announced publicly. The new protocol is designed to address many of the problems found in the older Wi-Fi protocols.  One of the primary features is an improved connection handshake protocol known as DragonFly, which is designed to address the known security vulnerabilities inherent in 802.11i.  Vanhoef & Piessens (2017) demonstrates a design flaw in the mutual authentication and session key agreement protocol that enables, under some scenarios, the ability for an attacker to decrypt and intercept traffic. Although WPA3 is designed to address this issue, a thorough analysis by Vanhoef and Ronen (2020) still finds numerous vulnerabilities in the newer handshake protocol. The personal implementation of WPA3 is still quite vulnerable in many of the same ways, although this issue has been documented and vendor patches were forthcoming as of April 2019.

A parallel capability, not included in the WPA3 spec, Opportunistic Wireless Encryption [OWE] will enable encryption of open networks. There aren't any specific known documented issues with OWE, however, as noted in Vanhoef's 2017 Black Hat paper (Vanhoef, 2017), if this approach or others utilizes broken handshake protocols, they will be inherently insecure. WPA3 and the newer protocols represent improvements in the protocol designs for Wi-Fi security, however from an end-to-end perspective, it is not clear that the underlying security issues from legacy protocols have been mitigated.

Ultimately, an assessment of the literature around the various 802.11 security implementations finds that only the most expensive, i.e. Geo-Fenced and EAP-TLS implementations, provide a high-degree of end-to-end security for users. Depending upon the version of 802.11 and which security implementation is used, the risk and threat factors change.

## 3. Recommender Systems and Automation

The concept of combining Automation and Recommender Systems to inform on-the-fly decision-making is not a new one. This technique underlies a number of applications in other domains that enable making decisions about selecting products and services. Resnick and Varian (1997) defines Recommender Systems as systems, not specific to the security domain, that can help make decisions based on recommendations from others. In a study on information security compliance, Bulgurcu, Cavusoglu and Bensabat (2010) finds informing users and providing awareness of negative impacts associated with not following desired security behaviors positively affects compliance. A mechanism that informs users and directs them to a best-practice-based outcome can be successful by providing the necessary information to drive best-practices compliant behaviors. Markotten and Gerd (2002) provide precedence for the concept of enabling users to manage their security through security awareness and providing appropriate choices through a well-designed and integrated user interface. An effective tool will provide the user with enough information and context to make the appropriate choice for their situation.

As Recommender Systems are defined, Adomavicius & Tuzhilin (2005) speak to content-based systems that use a Decision Tree Model approach. Based on an understanding of the problem, this is the most appropriate mechanism, given the information and context available. For a Recommender System to prove effective in the problem space of protecting users from dangerously configured W-Fi Access Points, it seems that it should present the decision space to the user with context. Needed context includes key information aligned each decision, such as relevant attributes of the access point, the connection, and other relevant meta-information. For a wireless access point, this will include physical location and information relative to determining the access point's authenticity and potential threat level. It is not reasonable to expect that the end user will ascertain the security of the system and then be inclined to document it in a recommendation. This will have to be done using an automated tool with some predefined criteria. Färber, Weitzel and Keim (2003) provide precedence for this approach and demonstrate a content-based model approach to filtering out personnel recruitment candidates. While the problem domain is different, conceptually there is similarity in the problem and approach. While the authors are filtering down from many candidates to an optimized list, the goal of this paper is to define a mechanism for filtering out high threat Wi-Fi access points, which presents a different scale to the challenge and approach. An attractive approach is to apply Collaborative Filtering, i.e. utilizing information from alike systems to amplify the effectiveness and speed of filtering (Thung, Lo, & Lawall,2013). The key limitation for this approach is that it will require more information than can expected to be immediately available and will require a unique repository to support the decision-making process. This may be a future area for study and improvement.

## 4. Analysis

Two considerations in developing an approach to the problem of informing risk-benefit decisions on utilizing specific wireless access points. The first consideration is whether there is sufficient information and domain knowledge to provide reliably good recommendations and the second is the practical application of such a tool. In terms of practicality, similar approaches have been utilized for other alike problems in the security domain and in other domains where crowd sourced and sensor data are integrated to apply decision advantage. One such area is in filtering Unsolicited Commercial Email [UCE], also known as spam. In this domain, Damiani et al. (2004), Kong et al. (2006), and Li, Zhong and Ramaswamy (2008) discussed collaborative approaches to filtering UCE. Damiani et al. (2004) integrate automated tagging with user input, which is very similar to the approach of tagging access points both with available data and telemetry, but also with user feedback. This approach has also been applied to problem sets such as traffic routing, product recommendations, and contractor vendor selection in the commercial mobile application space.

## 4.1. Collaborative Filtering

As an approach, Collaborative Filtering Systems is intended to address information overload with a few key considerations including intended end-user tasks, data sets and their properties (synthesized versus natural, biases, etc.) and accuracy metrics (Herlocker et al., 2004). In the vein of this specific problem, we have a clear set of user goals and user actions. The user's goal is to avoid malicious wireless access points. The user must make a decision whether or not to utilize a specific access point and to provide feedback on the access point, if it is used, both in terms of device telemetry and meta data, but also with direct observation.

The data set is a mixture between natural data, pulled from existing sources, i.e. WIGLE, as well as synthesized data generated on the fly by the tool and device telemetry. In terms of decision metrics, the criteria elicited about the access device should be flagged as critical negative, negative, potential negative and undetermined and presented to the user via the user-interface for decision making. This approach was chosen as an initial prototyping mechanism. Outside the scope of this work, a weighted criterion can be developed using the various attribute data associated with an access point and the aforementioned flagging levels to facilitate automating the decision-making based upon pre-configured user risk acceptance posture.

## 4.2. Design Issues

In terms of implementation, a key factor for success is properly architecting the system. Considerations in this regard include, but are not limited to the following concerns: the minimum required capability of the endpoint, the data/telemetry that must be collected from the endpoint, the degree to which the operating system platform of the endpoint allow for collecting said data, whether processing should take place on the device or upstream, or both, and which data is required for off-device processing. The work conducted in this assignment required significant re-architecting of the prototype application architecture implemented as a part of conducting this research as well as re-envisioning the Concept of Operations [CONOPS] for the tool.

Initially, the concept was to have the majority of the processing  take place on the device, which would include performing a local scan of wireless networks, filtering out prospective public access wireless access points, conducting basic analysis on them, allowing the user to connect to an access point, then doing more in-depth assessment.

Numerous challenges would render this CONOPS and architecture problematic. First, the chosen initial end point development platform (Android) presents some limitations. First, for this application, power and compute consumption are concerns. As noted by Kurihara et al (2017), applications that use the LocationManager functionality, as this application must use more power. Bao et al. (2016) noted that applications that are connectivity-focused have the most bug fixes and changes, many of which are directly related to device power consumption. Additionally, there is very limited selection of development plug-ins, and the event-driven nature of the platform for user applications seems likely to render a non-optimal user experience. The API platform is somewhat volatile in terms of deprecating functions and other changes over time that will affect the maintainability of the code.

Linares-Vasquez et al. (2015) found that API changes that break apps are an issue on the Android platform. This is important as some of the key functions rely on location information and access to network hardware, which are areas that are prone for change, both in how permissions work, and in how the actual API operates in the future. This application relies on soon-to-be deprecated functions for key capabilities. As a result, the more logical solution is to offload as much of the processing as possible to a more stable platform with more compute capacity and less platform limitation and volatility. All of the major processing can be done in a medium Amazon Lightsail instance, or similar compute and storage option, and made accessible via REST-API. The change in architecture allows for a much faster development cycle.

## 4.3.  Implementation Approach

The first critical path function for this application to work is the ability to scan for available wireless access points and to gather their Service Set Identifier [SSID], Basic Service Set Identifier [BSSID] – which equates to the Media Access Control [MAC] address, as well as the required encryption level and signal strength. The rationale for capturing this data follows. For known access points that exist in the WIGLE database containing data collected over time, it is possible to do a quick historical analysis – i.e. if the basic characteristics of the access point have changed. Additionally, as a full analysis, certain MAC addresses may potentially identify a suspicious access point. Specific characteristics related to the Organizationally Unique Identifier [OUI] can help to vet a device. If the OUI points to hardware that is known to be utilized for malicious purposes or is rarely used as a legitimate wireless access point, this can help to filter out threats. Of course, this is limited in effectiveness as an attacker can modify the default MAC address.

The application is operating under the assumption that public wireless access points will have public SSIDs and no encryption, although the intent is to make these user-toggle options in later revision of the prototype tool. If these attributes are spoofed, as in the Evil Twin Attack (Lanze et. Al, 2014), this data will not be sufficient to identify the malicious access point. It may be possible to discern malicious access points with a deeper scan, i.e. following the model of NMAP and fingerprinting the OS version of the device actively. This could be at least somewhat effective if it adds a temporal component, such as storing signatures over time and looking for deviations. Unfortunately, this is not possible on the Android platform, as this requires network access on the mobile device that is not possible without root privileges. As noted by Allen (2008), packet manipulation above the basic connect scan require higher-level privileges. Android heavily mirrors Linux and is further limited by the exposed abstractions and access to OS level primitives via the Android SDK.

On the Android platform, it is notable that the ability to generate a wireless scan is deprecated in future versions of the API (Google, n.d.). Because the current OS release, Android Pie, supports multiple API versions, the application was able to work around this. Based on the available documentation, at some point in the future, it may not be possible to have OS level access to request a wireless scan. As BSSIDs have become used for location purposes, there are privacy related issues in scanning and storing them. The scan returns the minimal level of information required to move to filtering and processing access points.

Accessing web services was accomplished utilizing an Android native library known as Retrofit. This library enables an event-triggered call and response interaction with Web Services. For each function call, JSON objects are exchanged over HTTPS between the Amazon Web Services Linux instance running a Flask Web application on the uWSGI platform behind a NginX web server front end. After quick prototyping with Android Studio, PyCharm and Postman multiple API calls were successfully implemented allowing for back end processing to take place. The current platform is able to provide analysis of the initial Access Point characteristics, return baseline DNS responses, and the WIGLE BSSID lookup is partially implemented.

## 5. Synthesis

Conceptually, this approach to end-point protection from malicious access points seems viable. There are a few factors to be discussed that will affect the degree to which underlying assessment methodology can work in an operational context. In an operational context, there are two feedback mechanisms that will help to form the recommendation system input. The limitations of those inputs will determine the potential of this architecture.

The first input stream is based upon the data captured from the discovery of a wireless access point and the ability to use that data to derive intent or characteristics of the device to which it connects. The SSID, for all intents and purposes, serves as a cosmetic identifier. Many of the attacks associated with Wireless Access Points are based upon presenting an SSID that will entice a user to connect or that matches a known Access Point that the end user will trust by default. Without additional context, this information is meaningless. Adding in the BSSID does not appreciably improve upon this by itself. While it is possible and worthwhile to do some analysis on the OUI field of the BSSID, preliminary testing of the prototype application found that less than 50% of wireless access points identified, both public and private, used BSSIDs that could be looked up using Wire Shark and NMAP OUI tables. The most likely explanation is that the MAC addresses are being randomized. Numerous scholarly papers and technical security forums discuss this as an endpoint privacy enhancement, but further analysis will be required to authoritatively assess why such a high percentage of access points were not identifiable. The conclusion based on the most readily available data is that sorting wheat from chaff using this information alone is not possible.

Incorporating WIGLE data enhances the value of the information the application can gather. Incorporating WIGLE enables identifying if the SSID or BSSID is known and if the data have changed over time. Wigle also allows incorporating spatial location data, which adds much more value to the data, but also presents potential for privacy risk and requires more permission for the end point application. The challenge here is the viability of the data. In order for the application to provide valuable input, there will have to be sufficient number of installed users feeding information to WIGLE to keep the data current. The solution is to have the application provide updates to WIGLE, but this requires further privacy encroachment and limitations. Assuming that WIGLE data is current, it enables the scan data to approach the utility of a whitelist, depending upon the risk appetite of the end user.

Utilizing a whitelist approach to securing the end-point does not appreciably add to its security. This is a slightly more informed approach compared to having the end user choose their public wireless access points manually. In order to add value to the existing data, a device signature is needed for malicious access points or further context is needed to conduct analysis. The ability to generate a signature for malicious access points and capture that data from mobile devices before connecting to them is well beyond the scope of this activity, limited by the hardware and operating system platform capabilities and will need to be treated separately. This leaves mechanisms for capturing context about access points and post connection signature analysis as means of conducting useful analysis to reduce risk.

There are some major drawbacks to doing a post connection analysis of an access point to try to define risk. Essentially, having connected to the access point, any services that are using internet connectivity are at risk for having communications intercepted, hijacked or modified. The access point can use connection attempts as an opportunity to download malware to the device to exploit common vulnerabilities to install back doors, ex-filtrate private data and expose other accounts and services, as well as install ransom ware, etc. Further investigation is needed in this regard, as the degree to which the platform can allow or disallow connections while in this state, can make this a possibly viable approach provided that the risk to the user can be effectively managed. If this is possible, using data pulled from the back-end architecture, the device can validate DNS resolution, SSL connections to known sites, and other indicators to assess the connection. This can provide an assessment of the risk of the access point that is only available while the user's device is connected to it. There is still the possibility that the access point could trigger malicious activity on specific DNS names or IP ranges or applications that are not in the scope of the tool's analysis.

Finally, the last remaining mechanism for conducting analysis of the end point is based on user feedback over time. This is taking a similar approach to WaZe, the well-known web application that overlays contextual data entered by users and other sensors to assess traffic flow and route safety factors such as presence of objects in the roadway, accidents, and weather conditions. A similar approach in the vein of endpoint device security is to allow users to report on the access points they connect to and report statuses such as: unable to browse internet,  or whether applications work or not. Tagging this data in concert with WIGLE historical data and the scan results over time can provide some limited enhanced end point security over the bare operating system. Mitigating the threat is complex and is limited by numerous factors.

## 6. Conclusion

In conclusion, Recommender Systems and Collaborative Filtering are a potential approach that can be applied against the rogue Wireless Access Point problem from an end-user perspective. The problem presented involves gathering relevant data to inform a risk calculation to inform whether to use an untrusted and unknown public wireless access point. It has been shown that at least some portion of the information needed to inform such decisions can be derived from open sources and via on the fly analysis. There are challenges specific to each end point platform for building such a tool, including, but not limited to the available user-space access to the underlying hardware, user-permissions framework and API flexibility to implement such tools in user space. Whether these can be overcome is a unique to each device eco-system and platform. There is the potential for further conceptual research in this space. If the concept proves out, it may be more effectively implemented at the platform level, which would obviate the need to expose limited hardware and OS features though the API and to user space programs.

The prototype is currently in an unfinished concept state although it will be completed in a further iteration of this paper. There is still further research to be done to refine both the solution space, given the limitations of the various end-point platforms and the API access, and the overall approach. One issue uncovered in this study was the lack of strong indicator to more effectively and quantitatively identify risk associated with connecting to specific access points given their characteristics. Further profiling and study may address this issue.